\documentclass[letter]{aa}  

\usepackage{graphicx}
\usepackage[version=4]{mhchem}
\usepackage{txfonts}
\begin{document}

\title{If you like C/O variations, you should have put a ring on it}

\author{Nienke van der Marel 
\inst{1,2}
\and
Arthur Bosman \inst{3}
\and
Sebastiaan Krijt \inst{4}
\and
Gijs D. Mulders \inst{5,6}
\and
Jennifer B. Bergner \inst{7}
}

\institute{Physics \& Astronomy Department, 
University of Victoria, 
3800 Finnerty Road, 
Victoria, BC, V8P 5C2, 
Canada
\and
Banting Research fellow
\and
Department of Astronomy, University of Michigan, 323 West Hall, 1085 S. University Avenue, Ann Arbor, MI 48109, USA
\and
School of Physics and Astronomy, University of Exeter, Stocker Road, Exeter EX4 4QL, UK 
\and
Facultad de Ingenier\'ia y Ciencias, Universidad Adolfo Ib\'a\~nez, Av.\ Diagonal las Torres 2640, Pe\~nalol\'en, Santiago, Chile
\and
Millennium Institute for Astrophysics, Chile
\and
Department of Geophysical Sciences, University of Chicago, Chicago, IL 60637, USA
}

\date{Received July 16th, 2021; accepted August 16th, 2021}

\abstract
{The C/O-ratio as traced with C$_2$H emission in protoplanetary disks is fundamental for constraining the formation mechanisms of exoplanets and our understanding of volatile depletion in disks, but current C$_2$H observations show an apparent bimodal distribution which is not well understood, indicating that the C/O distribution is not described by a simple radial dependence.} 
{The transport of icy pebbles has been suggested to alter the local elemental abundances in protoplanetary disks, through settling, drift and trapping in pressure bumps resulting in a depletion of volatiles in the surface and an increase of the elemental C/O.} 
{We combine all disks with spatially resolved ALMA C$_2$H observations with high-resolution continuum images and constraints on the CO snowline to determine if the C$_2$H emission is indeed related to the location of the icy pebbles.} 
{We report a possible correlation between the presence of a significant CO-icy dust reservoir and high C$_2$H emission, which is only found in disks with dust rings outside the CO snowline. In contrast, compact dust disks (without pressure bumps) and warm transition disks (with their dust ring inside the CO snowline) are not detected in C$_2$H, suggesting that such disks may never have contained a significant CO ice reservoir.} 
{This correlation provides evidence for the regulation of the C/O profile by the complex interplay of CO snowline and pressure bump locations in the disk. These results demonstrate the importance of including dust transport in chemical disk models, for a proper interpretation of exoplanet atmospheric compositions, and a better understanding of volatile depletion in disks, in particular the use of CO isotopologues to determine gas surface densities.}


\maketitle

\section{Introduction} 
\label{sec:intro}
Elemental abundance ratios such as C/O and C/H are some of the most fundamental properties of the composition of protoplanetary disks and, as a result, of gas giant exoplanetary atmospheres \citep[e.g.][]{Madhusudhan2019,Cridland2020}. The C/O-ratio in exoplanet atmospheres is considered a key measurement to link to the formation location of the planet in the disk. Whereas the classical view of an increasing step~function in the disk at snowline locations \citep{Oberg2011} has been expanded to a more advanced view due to pebble drift \citep{Oberg2016,Booth2017,Krijt2020,Schneider2021}, the precise mechanisms leading to variations in C/O are still poorly understood \citep[e.g.][]{Bosman2021}. Current disk surveys of C/O tracers such as C$_2$H \citep{Bergin2016,Semenov2018,Cleeves2018} appear to show a bimodal distribution of disks with and without C$_2$H detection indicating either high (C/O$>$1) or low (C/O$<$1) ratios \citep{Bergner2019,Miotello2019} but it is unclear whether this bimodality is an intrinsic property or the result of physical-chemical processes. The recent discovery of low C/O at the location of the asymmetric Oph~IRS~48 dust trap \citep{Booth2021-irs48} suggests that pressure bumps may play a key role in regulating the C/O profile in disks, which is particularly relevant considering that pressure bumps are usually associated with giant planets  \citep[e.g.][]{Pinilla2012a,vanderMarel2016-isot} and those planets would  accrete the composition of the gaseous material at the bump location. Pressure bumps act as dust traps, where large dust grains concentrate due to the gas-grain drag forces \citep[e.g.][]{Birnstiel2010}, which results in the appearance of dust rings in continuum images of protoplanetary disks \citep{Pinilla2012b}. In this manuscript the term dust ring and pressure bump are used interchangeably, although dust rings without pressure bumps may also exist \citep[e.g.][]{Okuzumi2016,Vericel2021}. The samples of the C$_2$H surveys by \citet{Bergner2019,Miotello2019} are biased towards brighter millimeter continuum disks, which are known to contain pressure bumps more often \citep{Pinilla2020,vanderMarelMulders2021}, as either ring disks  \citep[e.g.]{Andrews2018} or transition disks with inner cavities \citep[e.g.][]{Francis2020}. However, a clear correlation between C$_2$H emission and dust substructure has so far not been identified \citep{Bergin2016,Miotello2019,Bosman2021}.

A connection between volatile composition and pebble transport has been previously suggested for infrared line measurements of H$_2$, CO and H$_2$O in TW Hya \citep{Bosman2019}, and for the anti-correlation of the H$_2$O/HCN ratio of warm rovibrational lines and radial dust disk size \citep{Najita2013,Banzatti2020}. They pose a scenario where H$_2$O vapor is elevated for the most compact disks where dust has drifted inside the H2O snowline, whereas disks with pressure bumps show lower or no H$_2$O emission. The direct irradiation of the inner cavity wall may play an important role in particular in the desorption of oxygen-rich volatiles in transition disks \citep{Cleeves2011,vanderMarel2021-irs48}. Although dust rings are commonly detected in the brightest disks, they may actually not be that common in the total disk population \citep{vanderMarelMulders2021}: the selection bias of chemistry studies towards brighter disks could be a severe limitation in our understanding of the C/O in disks.

Interestingly, changes in the C/O-ratio as a result of pebble transport may also reflect changes in the CO/H$_2$, which is $\sim$few$\cdot10^{-4}$ in the interstellar medium (ISM). In models including pebble transport \citep{Oberg2016,Krijt2020}, molecular CO is depleted by 1-2 orders of magnitude in the outer disk, which is indeed seen in spatially resolved CO isotopologue observations \citep{Schwarz2016,Schwarz2019,Zhang2019} and shown to happen in the transition between the protostellar to the protoplanetary disk phase \citep{Zhang2020evolution,Bergner2020}. As the mixture of ices that are most affected by this depletion process (e.g. H$_2$O, CO$_2$, CO) is overall oxygen-rich, this process can result in outer disk surface layers with C/O$>$1 in the gas-phase \citep{Krijt2020,Bosman2021}.

CO depletion can explain why CO isotopologue measurements of protoplanetary disks appear to result in very low gas masses \citep{Ansdell2016,Miotello2017,Long2017}, in particular in comparison with alternative gas measurements such as HD \citep{Bergin2013,McClure2016} and typical disk accretion rates \citep{Manara2016}. As the gas distribution in protoplanetary disks is fundamental for models of e.g. giant planet formation, disk evolution and planet-disk interaction  \citep[e.g.][]{Helled2014,Andrews2020}, a better understanding on the effect of dust substructure on the volatile composition would also inform disk models used to infer gas masses from CO observations.

In this letter we aim to study the connection between the C/O-ratio in the gas in the outer disk, dust substructure and pebble transport by a detailed analysis of the spatially resolved C$_2$H profiles from the literature and the highest resolution ALMA continuum images available for these sources. We will focus in particular on the position of the CO snowline w.r.t. the dust rings, as dust pebbles rich in CO ice are expected to drift inwards to the nearest pressure bump and release the CO into the gas-phase inside the CO snowline through sublimation, resulting in an enrichment of the CO abundance and restoring the C/O-ratio to $\sim$1. However, dust rings may be present either inside or outside the CO snowline, depending on their location and the brightness of the star. The aim of this study is to test if the combination of the stellar properties (and resulting CO snowline) and dust substructure regulates the radial C/O profiles in the disk.

\section{Observations}
\label{sec:observations}
Our sample consists of all protoplanetary disks with deep, spatially resolved C$_2$H observations from the literature, primarily from ALMA \citep{Bergin2016,Bergner2019,Miotello2019,Facchini2021} and unpublished ALMA data of  IRS~48. Most of these disks are observed in the 262.006 GHz C$_2$H hyperfine doublet ($N=3-2,J=7/2-5/2,F=4-3$ and $F=3-2$), and a handful in the 349.4 GHz doublet ($N=4-3,J=7/2-5/2,F=4-3$ and $F=3-2$). The Miotello sample was selected from the brightest continuum disks in Lupus with a CO isotopologue detection and the Bergner sample from bright continuum disks covering a range of stellar and age properties. We add three other disks: TW~Hya \citep{Bergin2016}, PDS~70 \citep{Facchini2021} and IRS~48 \citep{vanderMarel2021-irs48}. For DM~Tau, two published datasets exist \citep{Bergin2016,Bergner2019} and we choose to include the latter. 

The disks are observed at a range of sensitivities and are located at distances between 60 and 175 pc, covering a range of stellar luminosity, spectral type and age. The spatial resolution varies between 0.2" and 0.8", or $\sim$40 au spatial scales. The sample consists of 26 disks, with 20 C$_2$H detections and 6 non-detections. The sample is listed in Table \ref{tab:sample}. 

For all disks, high-resolution continuum data at either 870 $\mu$m or 1.3 mm (ALMA Band 7 or 6) have been taken with constraints on the dust substructure. Most continuum images have previously been published (references in Table \ref{tab:sample}) and have a typical spatial resolution of 0.05" or $\sim$8 au. The images are shown in Figure \ref{fig:gallery} and cover a range of dust morphologies, including compact, ring and transition disks. Compact disks are defined as disks without large scale gaps, following \citet{vanderMarelMulders2021}, transition disks as disks with a cleared inner dust cavity, and ring disks as disks with one or more gaps in the outer disk.

\section{Analysis}
\label{sec:analysis}
In order to investigate the role of dust substructure onto the presence of C$_2$H, the C$_2$H emission needs to be normalized spatially across the sample, as the sample consists of disks of a range of sizes, and the integrated flux (as used by the original studies) may not be representative. Therefore we use the azimuthally averaged radial profiles of the integrated C$_2$H emission, and use the peak value as our main measurement of the C$_2$H brightness, or the 3$\sigma$ upper limit of the integrated emission in case of a non-detection. As the doublet lines are blended, the integrated emission represents the integration over both lines. For the data in \citet{Bergin2016},  \citet{Bergner2019} and \citet{Facchini2021} such radial profiles are provided. For the sample of \citet{Miotello2019}, the azimuthally averaged profiles are computed from their zero-moment maps. All profiles are presented in Appendix \ref{sec:figures}. For IRS~48, the C$_2$H 349.4 GHz line upper limit was derived in Appendix \ref{sec:irs48}. In order to normalize the sample across spatial scales and beam sizes, the integrated peak flux density is converted to the brightness temperature and scaled to a distance of 100 pc. The resulting scaled peak brightness temperatures are listed in Table \ref{tab:sample}. 

The dust continuum images can provide a measurement of the amount of dust~mass inside and outside the CO snowline. The latter provides a measurement of the amount of CO-ice-covered dust~mass $M_{\rm dust,CO}$ or cold dust reservoir, which is hypothesized to dominate the regulation of the elemental abundance ratio. The CO snowline is defined as the location where the temperature (computed from the stellar luminosity) drops below 22 or 30 K and the details of this calculation are provided in Section \ref{sec:dustcalc}. The $M_{\rm dust,CO}$ values are listed in Table \ref{tab:sample}.

\begin{figure}[!ht]
    \centering
    \includegraphics[width=0.49\textwidth]{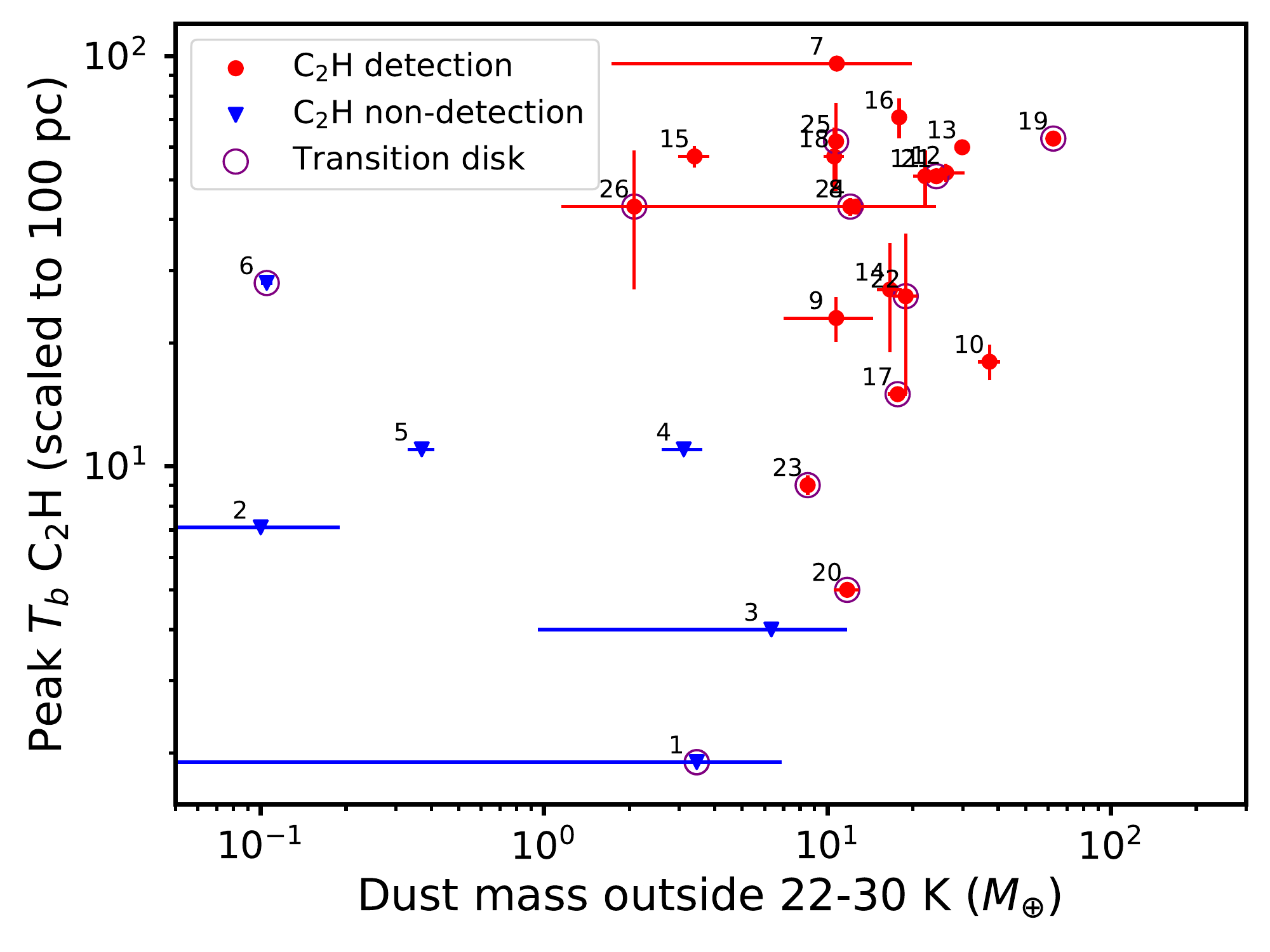}
    \includegraphics[width=0.49\textwidth]{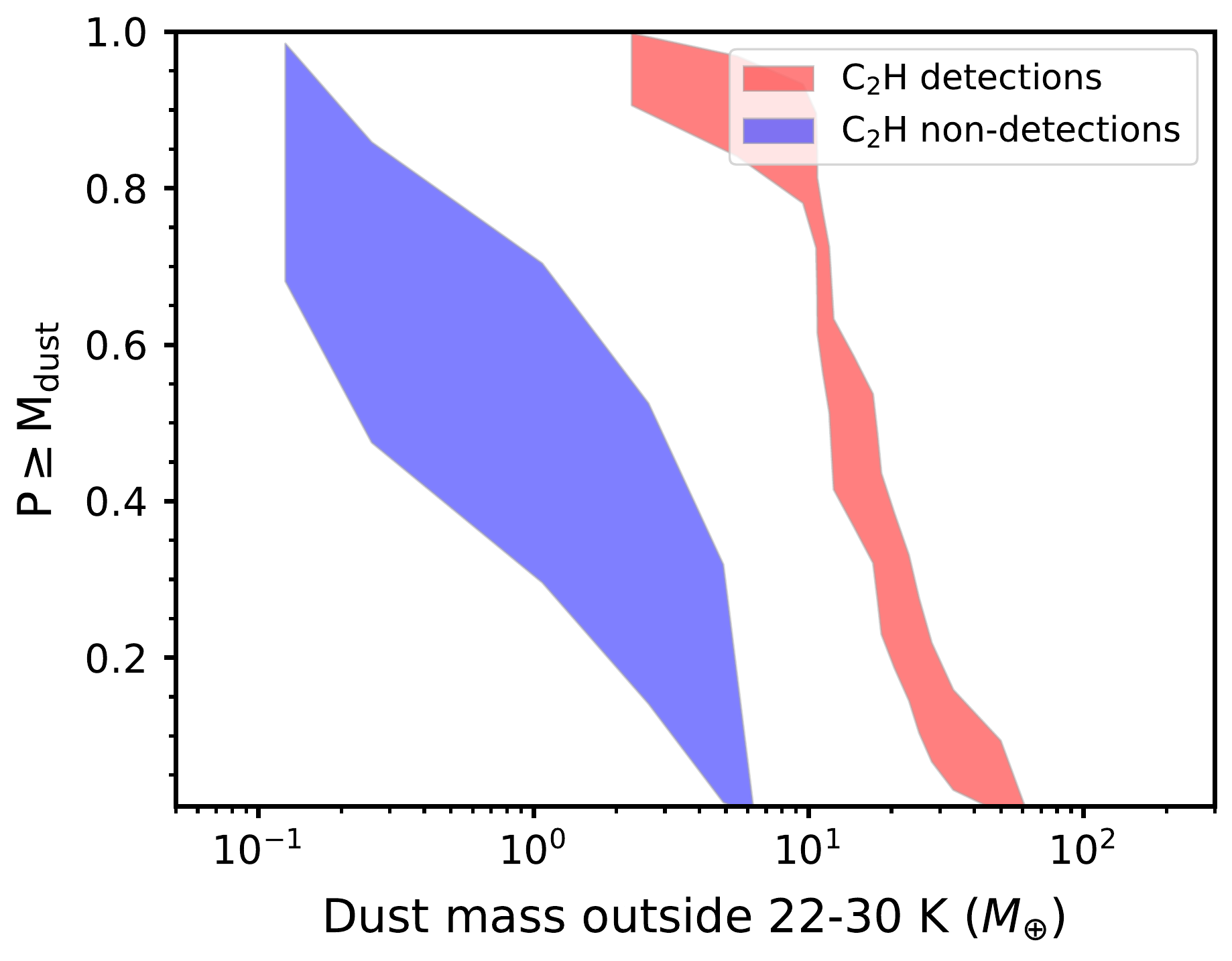}
    \caption{Comparison between the scaled peak flux of the C$_2$H radial profiles and the millimeter-dust~mass outside the CO snowline. The C$_2$H non-detections are indicated in blue, and these data points represent 3$\sigma$ upper limits. The C$_2$H detections are shown in red with 1$\sigma$ error bars. Transition disks are encircled. The numbers correspond to the targets as listed in Table \ref{tab:sample}. Horizontally, the error bars indicate the range of dust~masses for the chosen temperature of the CO freezeout temperature (22 to 30 K), where the datapoint represents the average between these two values. The bottom panel shows the cumulative distribution function of the dust~masses for the two groups, computed using the Kaplan Meier estimator (see text). The two-sample tests indicate that although there is some overlap between the two samples, they are statistically distinct and the C$_2$H non-detection is possibly linked to the lower amount of CO-ice covered millimeter-dust in the disk.}
    \label{fig:outerdustmass}
\end{figure}

\section{Results}
\label{sec:results}

\begin{figure*}[!ht]
    \centering
    \includegraphics[width=\textwidth]{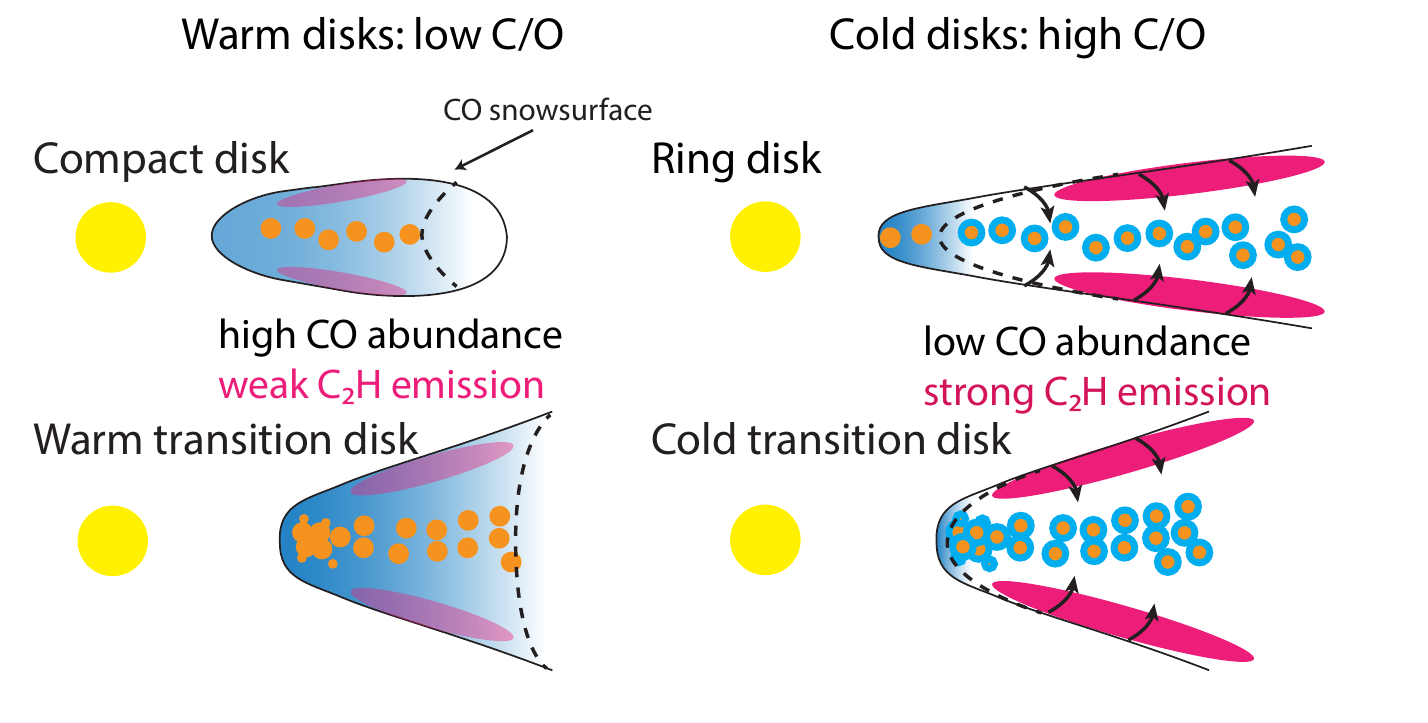}
\caption{Cartoon of the effects of dust transport in disks for different types of disks. Grain growth, settling and vertical mixing drain volatiles from the outer disk atmosphere and lock them in grains in the midplane leading to changes in C/O depending on the relative location of snowline and pressure bump. The C$_2$H emitting region is indicated, where purple indicates weak emission and magenta strong emission.
{\bf Left:} A warm disk with low C/O (low C$_2$H) could be either a compact dust disk without pressure bumps (hence efficient radial drift) or a transition disk with a dust ring (pressure bump) inside the CO snowline. In both cases the millimeter dust pebbles are not covered with CO ice.
{\bf Right:} A cold disk with high C/O (high C$_2$H) could be either a ring disk or a transition disk with a dust ring (pressure bump) outside the CO snowline, so there is a large CO-icy dust reservoir, which depletes the surface layers from volatiles.
}
    \label{fig:cartoon}
\end{figure*}

The disks with a C$_2$H non-detection have little or no dust emission outside the CO snowline (Figure \ref{fig:profiles}), whereas all C$_2$H detections still have a significant cold dust reservoir. The CO snowline as determined by a temperature profile is  a reasonable approximation, with the note that the measured CO snowline appears to lie further out for the transition disks (suggesting a somewhat warmer temperature profile) and somewhat closer in for the ring disks, where available. Due to the lack of CO snowline measurements for the bulk of the disks the dust~mass outside the CO snowline can only be derived consistently throughout the sample using the temperature profile, with the caveat that the dust~mass may be somewhat underestimated for the ring disks and overestimated for the transition disks. For the transition disks most dust rings lie at or outside the CO snowline, with the exception of IRS~48 and V4046~Sgr. For J16070854 the dust ring is difficult to identify due to the spatial resolution of the image, but visibility modeling indicates an inner cavity of $\sim$40 au \citep{vanderMarel2018}.

The scaled C$_2$H peak brightness varies across the sample from $<$1.1 up to 97 K~km~s$^{-1}$. The upper limits lie on the lower end of the detection values, but there is no obvious bimodal distribution as seen in the integrated flux \citep{Miotello2019}. The scaled C$_2$H peak temperature for PDS~70, V4046~Sgr, TW~Hya and IM~Lup is at a similar level as the upper limits.

Using these properties, we now look for a dependence of the C$_2$H peak flux on the amount of icy dust~mass in Figure \ref{fig:outerdustmass}. The data points show the range of dust~masses consistent with a CO snowline temperature of 22 to 30 K. As the dust emission is usually ring-like, the shift in CO snowline temperature has a profound effect in some dust~masses, and negligible in others, as can be seen also in the $M_{\rm dust,CO}$ values in Table \ref{tab:sample}. Beam dilution is possibly affecting the estimates for the lower resolution images so some of these values may be upper limits. 
Figure \ref{fig:outerdustmass} also shows the cumulative distributions of the disks with and without C$_2$H detection, using the Kaplan-Meier estimator (KME) from the ASURV package \citep{Lavalley1992}. The mean icy dust~mass of the C$_2$H-detection group is 18$\pm$3 $M_{\oplus}$, whereas the mean icy dust~mass of the C$_2$H-non-detection group is 2.0$\pm$0.9 $M_{\oplus}$. 

For the scaled C$_2$H peak fluxes we compute a two sample test using ASURV (to properly treat the upper limits of the C$_2$H data points). The Gehan's Wilcoxon test (to include upper limits) indicates a probability of 0.5\% that the two groups have been drawn from the same population. For the icy dust~masses, this probability is 0.06\%, so we can indeed conclude the groups are statistically distinct. In order to check that this is not caused by an underlying correlation with another parameter, we perform a Kolmogorov-Smirnov statistic test for stellar luminosity $L_*$ and total dust~mass $M_{\rm dust}$, and find a probability of 92\% for $L_*$ and a probability of 23\% for $M_{\rm dust}$ that the two groups have been drawn from the same distribution, so the difference is likely not caused by either of these properties. For the disk dust size (defined as the radius $R_{68}$ which encircles 68\% of the total flux), this probability is 3.6\%, so there could be an underlying relation there as well, although not as strong as with the icy dust~mass, and the two parameters are obviously related. 

\section{Discussion and conclusions}
\label{sec:discussion}
The connection between the C$_2$H peak flux and the icy dust~mass outside the CO snowline from Figure \ref{fig:outerdustmass} indicates a direct link between the C/O-ratio and the presence of a CO-ice pebble reservoir. Such a reservoir requires mm-dust at large radii, hence prevention of radial drift through pressure bumps outside the snowline. According to \citet[][Figure 10]{Krijt2020}, this results in a low CO abundance and elevated C/O$\gtrsim$1 in the surface layers in the regime outside the CO snowline, where the high C/O can be enhanced to $\gtrsim$1.5 as needed for a strong C$_2$H detection \citep{Bergin2016}, by additional vertical mixing and destruction of carbonaceous grains \citep{Bosman2021}. Inside the CO snowline the CO abundance returns to the ISM value of 10$^{-4}$ and C$_2$H is below typical detection thresholds, which is consistent with the radial C$_2$H profiles (see Section \ref{sec:figures}). The radial extent of C$_2$H is larger than the dust extent as the outer gas disk has been depleted by settling and radial drift of icy dust grains. The observed lack of anti-correlation between C$^{18}$O and C$_2$H by \citet{Bergner2020} is not at odds with this scenario, as they state that the C$_2$H production will not increase further with CO depletion when the C/O-ratio is already elevated \citep{Cleeves2018}.

In contrast, the disks with a C$_2$H non-detection contain little or no CO-ice-covered mm-dust: in fact, 4 out of 6 of these disks are compact in continuum emission and do not show any signs of substructure. The exceptions to this rule are IRS~48 and J16070384. IRS~48 has a bright, asymmetric dust trap, but this trap is located \emph{inside} the CO snowline unlike the other transition disks, whereas J16070384 shows a large edge-on dust ring, but is overall very faint in mm-dust. 

If low C/O (C$_2$H non-detection) is indeed linked to the lack of CO ice, this suggests that these disks may never have contained a significant cold dust reservoir. This is indeed possible for the compact disks, under the assumption that the disks were warmer in the embedded phase and the pebbles have drifted in before CO could freeze out. Faint CO emission in disks was recently suggested to be caused by these disks being compact and warmer \citep{Miotello2021}. In that case, the scenario of ice-covered pebble drift to inside the CO snowline by \citet{Krijt2020} would not actually occur in disks. The detection of H$_2$CO and CH$_3$OH emission in IRS 48  \citep{vanderMarel2021-irs48} implies that this disk has either been cold enough to have CO-ice in the past, or has inherited H$_2$CO and CH$_3$OH from the dark cloud phase \citep{Drozdovskaya2014,Booth2021-hd100} as CH$_3$OH-ice sublimates at higher temperatures.

Other than IRS~48, the transition disks in this sample have their dust rings outside the CO snowline, implying that the bulk of the CO ice is not returned to the gas phase and they are essentially the same as ring disks. The CO snowline in transition disks may lie further out than our estimates, as revealed by the direct measurements of the CO snowline of e.g. DM~Tau and LkCa~15 \citep[][]{Qi2019}. However, considering their radial profiles, this will not significantly decrease their icy dust~mass to the level of the C$_2$H non-detections. 

The four disks with low C$_2$H (IM~Lup, TW~Hya, PDS~70 and V4046~Sgr) may have a lower C/O-ratio than the bulk of the C$_2$H detections, consistent with the derived value of C/O$\sim$0.8 from the detailed analysis of IM~Lup \citep{Cleeves2018}. For TW~Hya, \citet{Bergin2016} derived C/O$\gtrsim$1, which may still be lower than the value of C/O$>$1 of the majority of the Lupus disks derived by \citet{Miotello2019}. These four disks may fall in between the two extremes of the C/O~ratios in our sample. Interestingly, all but IM~Lup are transition disks. Transition disks are known to have irradiated cavity walls, with higher temperatures and different compositions of sublimated species \citep{Cleeves2011,Bosman2019}. 

Alternatively, there is a possible correlation between the dust disk radius and the C$_2$H brightness, which would suggest that the C$_2$H non-detections may just be caused by a smaller dust surface density area. Such a correlation could be understood if C$_2$H emission is primarily produced by fragmentation of large grains that are mixed vertically into the hot surface layers, which is indeed predicted as a contributing factor to C$_2$H \citep{Bosman2021}. However, such a relation would not be able to explain the low C/O in IRS~48. The current sample of available C$_2$H observations is not well-constructed to fully test our proposed scenario. Targeted, high-resolution observations of disks with pressure bumps inside and outside the CO snowline are needed.

In summary, this work suggests that the C/O value in disks (as traced by the C$_2$H emission) is related to the formation and transport of CO ice-covered dust pebbles. Figure \ref{fig:cartoon} presents a summary of the proposed scenario. In large, cold ring disks the existence of cold pebble reservoir results in a low CO abundance and uniform, high C/O ratio in the surface layers. \footnote{If some pebbles can filter through the pressure bumps, the direct vicinity of the CO snowline could see high gas-phase CO abundances and C/O$\sim$1.} Alternatively, compact and/or warm disks do not efficiently sequester CO-ice in the midplane and retain high CO and low C/O abundances at all radii. In general, the efficiency of CO ice sequestration in the midplane depends on the interplay between dust coagulation, vertical settling, and turbulent mixing.

In transition disks, irradiation of cavity walls increases the desorption of ices \citep[][Fig.~2b]{Cleeves2011}, possibly lowering the gaseous C/O-ratio to $\sim0.5$ further out, as the temperature gets high enough to desorb oxygen-rich ices. This means that the C/O-ratio of giant planets inside the gaps is set by the complex interplay between pressure bumps and snowlines. 
If this hypothesis is correct, chemical disk models need to take dust transport and pebble history into account when analyzing CO and other molecular lines for proper interpretation, with new potential to use CO~isotopologue data to derive gas surface densities in disks. Using the current dust morphology, the effect of CO depletion due to dust transport can be estimated directly and used to set the CO abundance in the disk, used to derive gas surface densities.

This result is particularly timely as many spectra of exoplanet atmospheres are expected to be delivered in the coming years with e.g. the James Webb Space Telescope (JWST). As the exoplanet atmospheric composition and C/O-ratio is set by accretion of gas and icy pebbles, including gas released from fragmentation of icy pebbles \citep{Cridland2020}, its formation history can only be interpreted when the C/O distribution in disks is properly understood.

\begin{acknowledgements}
We would like to thank the referee for their constructive comments which have helped to clarify the manuscript, and we thank Anna Miotello for useful feedback and Feng Long for the provided continuum data of the Taurus disks. 
N.M. acknowledges support from the Banting Postdoctoral Fellowships program, administered by the Government of Canada. G.D.M. acknowledges support from ANID --- Millennium Science Initiative ---  ICN12\_009. 
ALMA is a partnership of ESO (representing its member states), NSF (USA) and NINS (Japan), together with NRC (Canada) and NSC and ASIAA (Taiwan) and KASI (Republic of Korea), in cooperation with the Republic of Chile. The Joint ALMA Observatory is operated by ESO, AUI/ NRAO and NAOJ. This paper makes use of the following ALMA data: 2013.1.00198.S, 2013.1.00226, 2015.1.00671.S, 2015.1.00964.S, 2016.1.00627.S, 2016.1.00459.S, 2015.1.00888.S, 2017.A.00006.S, 2017.1.01167.S, 2017.1.00834.S, 2019.1.01619.S.  
\end{acknowledgements}

\vspace{5mm}

\appendix

\bibliographystyle{aa}

\appendix
\onecolumn

\section{Icy dust~mass calculation}
\label{sec:dustcalc}
\begin{figure*}[!ht]
    \centering
    \includegraphics[width=0.8\textwidth]{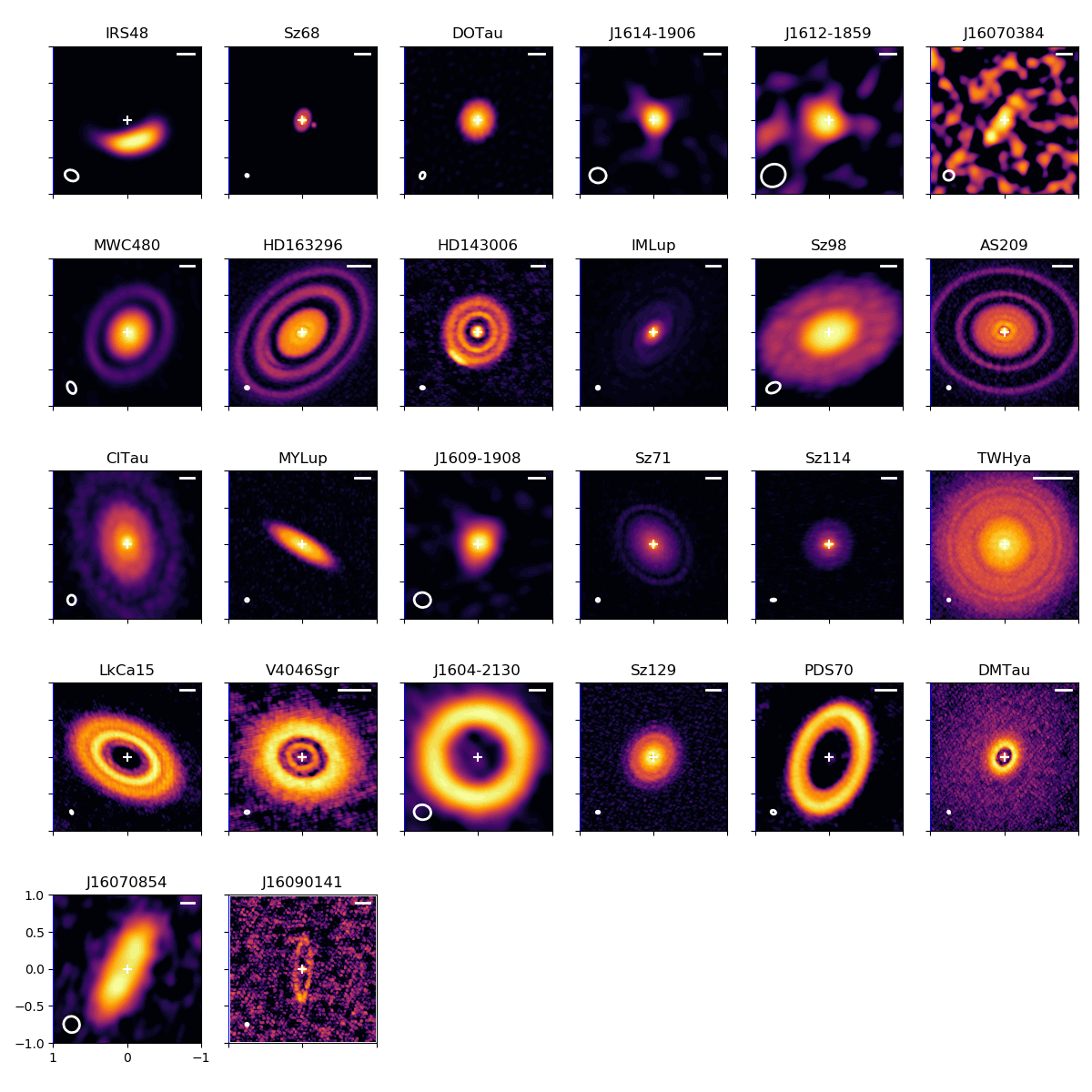}
    \caption{Continuum image gallery of the sources in our sample, shown with an asinh stretch. The beam size is indicated with a white ellipse in the lower left corner. A bar in the top right indicates the 30 au scale. The top row shows the sources without C$_2$H detection (mostly compact disks (CD)), followed by the disks with C$_2$H detection: first the ring disks (RD), followed by the transition disks (TD), sorted by decreasing stellar luminosity.}
    \label{fig:gallery}
\end{figure*}

For our analysis we need a measurement of the amount of dust~mass outside the CO snowline, the icy dust~mass, using the continuum images presented in Figure \ref{fig:gallery}.
The CO snowline is known to be located at distances corresponding to a range of temperature values between 18 and 25 K based on observations \citep{Pinte2018imlup,Qi2019} and based on laboratory work, is expected to be around 22 K for pure CO ice \citep{Bisschop2006} or as high as 30 K for CO mixed in H$_2$O ice \citep{Noble2012}. Exact snowline locations in disks have been attempted to be derived using snowline tracers such as N$_2$H$^+$ and DCO$^+$ but detailed 2D models have shown that their emission is more complex than a simple anti-correlation with CO \citep{Qi2015,vantHoff2017}. Therefore, we will adopt the 22 K and 30 K temperatures as the two extremes of the CO snowline location. For the temperature profile we adept the simplified expression of the midplane temperature in a passively heated, flared disk in radiative equilibrium \citep{Chiang1997,Dullemond2001}:
\begin{equation}
\label{eqn:temperature}
T(r) = \Big(\frac{\phi L_*}{8\pi\sigma_Br^2}\Big)^{1/4} 
\end{equation}
with $\sigma_B$ the Stefan-Boltzmann constant, $\phi$ the flaring angle (taken as 0.02) and $L_*$ the stellar luminosity. Using this equation, we can deduce the radius $r_{\rm CO}$ where $T$=30/22 K in the midplane. For many of the disks in our sample the CO snowline has been derived directly from the line data: \citet{Zhang2019} derived the snowlines of DM~Tau, TW~Hya, IM~Lup and HD163296 at $\sim$12, 20, 30 and 60 au, respectively, which is consistently further in than the 22 K line from the temperature profile from Eqn. \ref{eqn:temperature}. \citet{Qi2019} derived the CO snowline using detailed modeling of the N$_2$H$^+$ emission profile for DM~Tau, LkCa~15 and GM~Aur at 75$^{+10}_{-30}$, 58$^{+6}_{-10}$ and 48$^{+10}_{-8}$ au, respectively, and set a limit of $<$33 au for the V4046~Sgr system, the discrepancy for DM~Tau may be due to the complex relation between N$_2$H$^+$ and the snowline \citep{vantHoff2017}. These values are all located further out than our 22 K line, which is expected as all these disks have large inner cavities where the temperature structure is altered due to the irradiation of the cavity wall \citep{Cleeves2011}. For the remaining targets, no CO snowline measurements are available in the literature.

\begin{table*}[!ht]
    \centering
    \caption{Sample properties}
    \label{tab:sample}
    \footnotesize
    \begin{tabular}{l|ll|lll|llllll|l}
    \hline
     Target & \multicolumn{2}{c|}{Stellar}&\multicolumn{3}{c|}{C$_2$H line properties}&\multicolumn{6}{c}{Continuum properties}\\
     \hline
     & $L_*$ & $d$ & $F_{integ}$ & $T_{b,peak,100pc}$ & Beam & $F_{total}$ & $\nu$ & rms & $M_{\rm d<30K}$ & $M_{\rm d<22K}$ & Beam & Refs$^b$\\
    &($L_{\odot}$)&(pc)&(mJy&(K &(")&(mJy)&(GHz)&(mJy &($M_{\oplus}$)&($M_{\oplus}$)&(")&\\
    &&&km s$^{-1}$)&km s$^{-1}$)&&&&bm$^{-1}$)&&\\
    \hline
1. IRS48&18&135&$<$11$^a$&$<$1.1$^a$&0.63x0.50&173&336&0.05&4.24&0.01&0.19x0.14&1,6\\
2. Sz68&5.5&154&$<$5.3&$<$7.1&0.25x0.21&76&239&0.02&0.19&0.01&0.04x0.03&2,7\\ 
3. DOTau&1.4&139&$<$62&$<$2.4&0.63x0.49&120&225&0.10&11.77&0.95&0.09x0.07&3,8\\ 
4. J1614-1906&0.5&143&$<$100&$<$5.3&0.56x0.45&20&239&0.04&3.62&2.60&0.22x0.20&3,9\\ 
5. J1612-1859&0.3&139&$<$97&$<$3.6&0.56x0.45&5.7&341&0.26&0.42&0.33&0.33x0.30&3,10\\ 
6. J16070384&0.25&150&$<$16&$<$24&0.25x0.21&0.5&225&0.04&0.11&0.10&0.13x0.12&2,11\\ 
7. MWC480&18.6&162&1846&97&0.73x0.47&264&225&0.06&20.0&1.7&0.17x0.11&3,8\\
8. HD163296&17&101&4396&41&0.55x0.47&634&239&0.02&24.1&1.1&0.05x0.04&3,7\\
9. HD143006&3.8&165&435&20&0.58x0.46&57&239&0.01&14.6&7.0&0.05x0.04&3,7\\
10. IMLup&2.6&158&1450&18&0.56x0.55&209&239&0.01&40.8&34.0&0.04x0.04&3,7\\
11. Sz98&1.53&156&1101&51&0.22x0.17&104&255&0.08&24.3&20.0&0.20x0.13&2,2\\
12. AS209&1.4&121&2709&52&0.65x0.55&233&239&0.02&30.6&21.9&0.04x0.04&3,7\\
13. CITau&0.9&159&1041&59&0.59x0.49&128&225&0.06&31.6&28.2&0.13x0.11&3,8\\
14. MYLup&0.87&157&284&27&0.22x0.17&78&239&0.02&18.4&14.9&0.04x0.04&2,7\\
15. J1609-1908&0.4&138&563&56&0.56x0.45&22&239&0.04&3.82&2.98&0.22x0.20&3,9\\ 
16. Sz71&0.33&156&1101&71&0.23x0.17&83&239&0.02&19.1&16.9&0.04x0.04&2,7\\
17. TWHya&0.3&60&10720$^a$&15$^a$&0.48x0.39&1380&346&0.03&19.2&16.3&0.03x0.03&4,12\\
18. Sz114&0.2&162&383&57&0.22x0.17&48&239&0.02&11.5&9.7&0.07x0.03&2,7\\
19. LkCa15&1.3&159&2435&62&0.59x0.49&229&341&0.02&24.1&24.1&0.05x0.03&3,13\\
20. V4046Sgr&0.86&72&2962&4.8&0.97x0.70&251&239&0.05&13.0&10.5&0.06x0.04&3,6\\
21. J1604-2130&0.7&150&2632&52&0.56x0.45&258&350&0.18&24.2&24.2&0.23x0.20&3,6\\
22. Sz129&0.44&162&107&26&0.26x0.21&82&239&0.02&20.9&17.0&0.04x0.03&2,7\\
23. PDS70&0.3&113&1327&8.7&0.39x0.32&159&351&0.03&8.5&8.5&0.07x0.05&5,7\\
24. DMTau&0.24&145&2013&43&0.57x0.49&56&225&0.01&12.0&12.0&0.03x0.02&3,7\\
25. J16070854&0.18&176&647&62&0.22x0.17&33&225&0.11&10.8&10.6&0.22x0.21&2,11\\
26. J16090141&0.15&164&200&37&0.25x0.21&7.6&225&0.02&2.1&2.1&0.04x0.03&2,14\\ 
    \hline     
    \end{tabular}\\
    $^a$ C$_2$H $N=4-3$,  $J=7/2-5/2$ doublet at 349.4 GHz rather than the $N=3-2$, $J=7/2-5/2$ doublet at 262.006 GHz. \\
    $^b$ References for the C$_2$H data (first) and for the continuum data (second).\\
    {\bf Refs.} 1) This work; 2) \citet{Miotello2019}; 3) \citet{Bergner2019}; 4) \citet{Bergin2016}; 5) \citet{Facchini2021}; 6) \citet{Francis2020}; 7) \citet{Andrews2018}; 8) \citet{Long2019}; 9) ALMA archival data 2017.1.01167.S; 10) \citet{Barenfeld2016}; 11) \citet{Ansdell2018}; 12) \citet{Andrews2016}; 13) \citet{Facchini2020}; 14) van der Marel et al. in prep.
\end{table*}

\begin{figure*}[!ht]
    \centering
    \includegraphics[width=\textwidth]{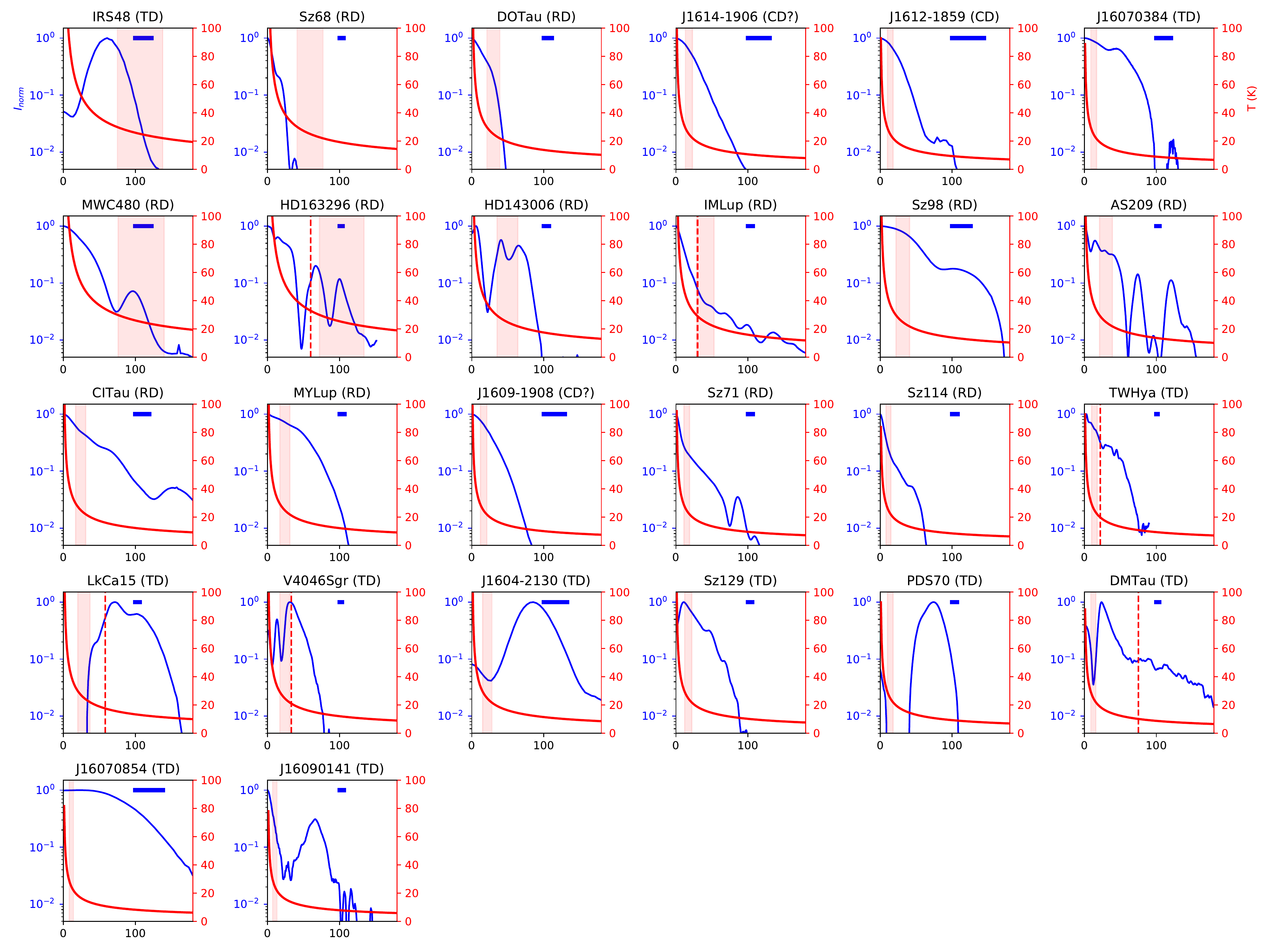}
    \caption{Azimuthally averaged and normalized intensity profiles of the dust continuum (blue) in combination with the temperature as computed from the stellar luminosity using Eqn. \ref{eqn:temperature} (red). The transparent red bar indicates the 22-30 K temperature range (approximate CO snowline) according to this temperature profile. In some cases the CO snowline has been derived directly from observations (see text), which is indicated with a vertical dashed red line. A blue horizontal bar indicates the beam size. The top row shows the sources without C$_2$H detection (mostly compact disks (CD)), followed by the disks with C$_2$H detection: first the ring disks (RD), followed by the transition disks (TD), sorted by decreasing stellar luminosity.}
    \label{fig:profiles}
\end{figure*}

Figure \ref{fig:profiles} shows the normalized azimuthally averaged profiles of the dust continuum, with overlaid the temperature profile from Eqn. \ref{eqn:temperature} and the corresponding 22-30 K range (expected CO snowline location), plus the CO snowline derived directly, where available. 

The continuum flux is computed by integrating the flux density over all pixels$>3\sigma$ outside $r_{\rm CO}$, with $\sigma$ the rms noise. The corresponding dust~mass is then computed using the standard assumptions of isothermal, optically thin emission of 20 K as demonstrated by \citet{Hildebrand1983} and regularly used in ALMA disk continuum surveys \citep[e.g.][]{Ansdell2016}:
\begin{equation}
    M_{\text{dust }}=\frac{F_{\nu,CO} d^{2}}{\kappa_{\nu} B_{\nu}\left(T_{\text{dust}}\right)}
\end{equation}
where $B_{\nu}$ is the Planck function for a characteristic dust temperature, $T_{\text{dust}}$, $\kappa_{\nu}$ the dust grain opacity, $d$ the distance to the target in parsecs, and $F_{\nu,CO}$ the millimeter flux outside $r_{\rm CO}$. Due to the lower spatial resolution of J1609-1908, J1612-1859 and J1614-1906, these icy dust~masses should be considered upper limits.

\section{C$_2$H profiles}
\label{sec:figures}
This section contains the radial profiles of the C$_2$H emission from the zero-moment maps in the literature \citep{Bergin2016,Miotello2019,Bergner2019,Facchini2021}.

\begin{figure*}[!ht]
    \centering
    \includegraphics[width=\textwidth]{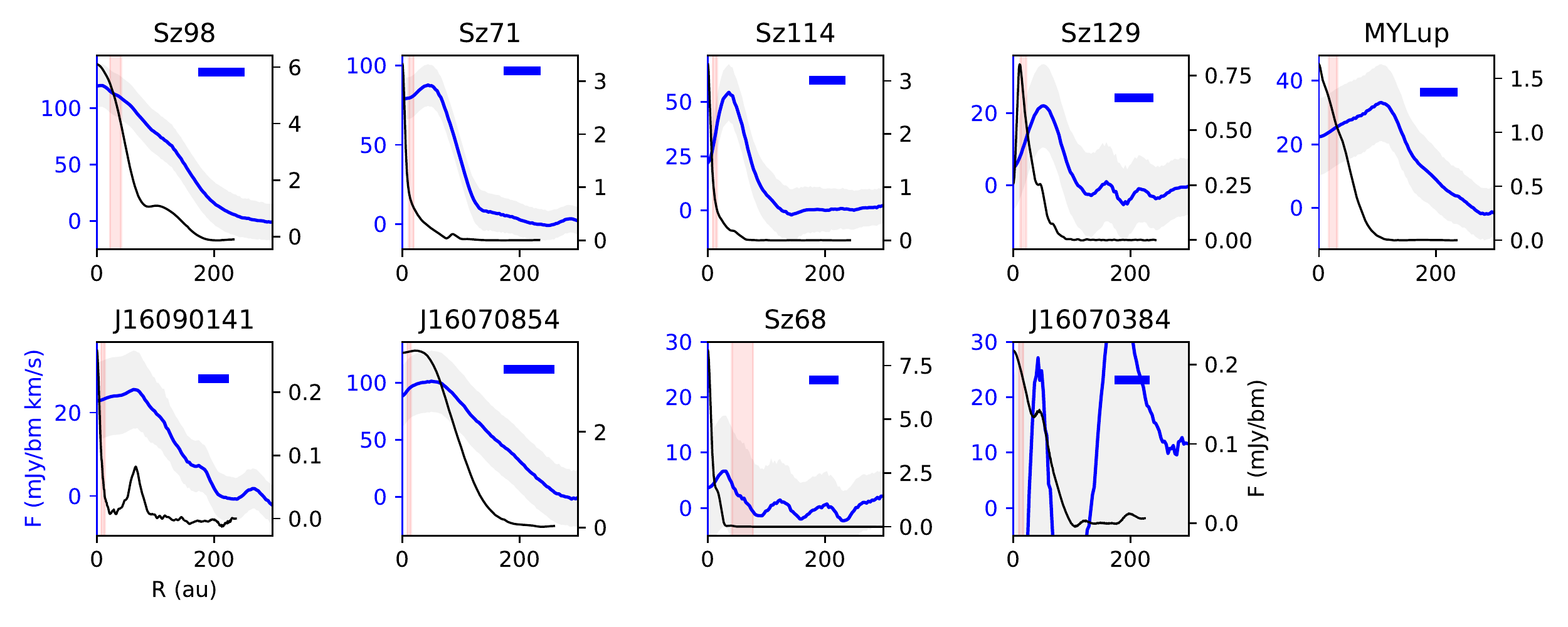}
    \caption{Radial C$_2$H profiles of the Miotello sample (blue). The grey indicates the uncertainty. The CO snowline (in the 22-30 K range) is overplotted as in Figure \ref{fig:profiles} and the beam size is indicated as a blue bar in the top right corner. The high-resolution continuum profiles of Figure \ref{fig:profiles} are overplotted in black to show that the C$_2$H is always more extended than the continuum.}
    \label{fig:c2h_miotello}
\end{figure*}

\begin{figure*}[!ht]
    \centering
    \includegraphics[width=0.9\textwidth]{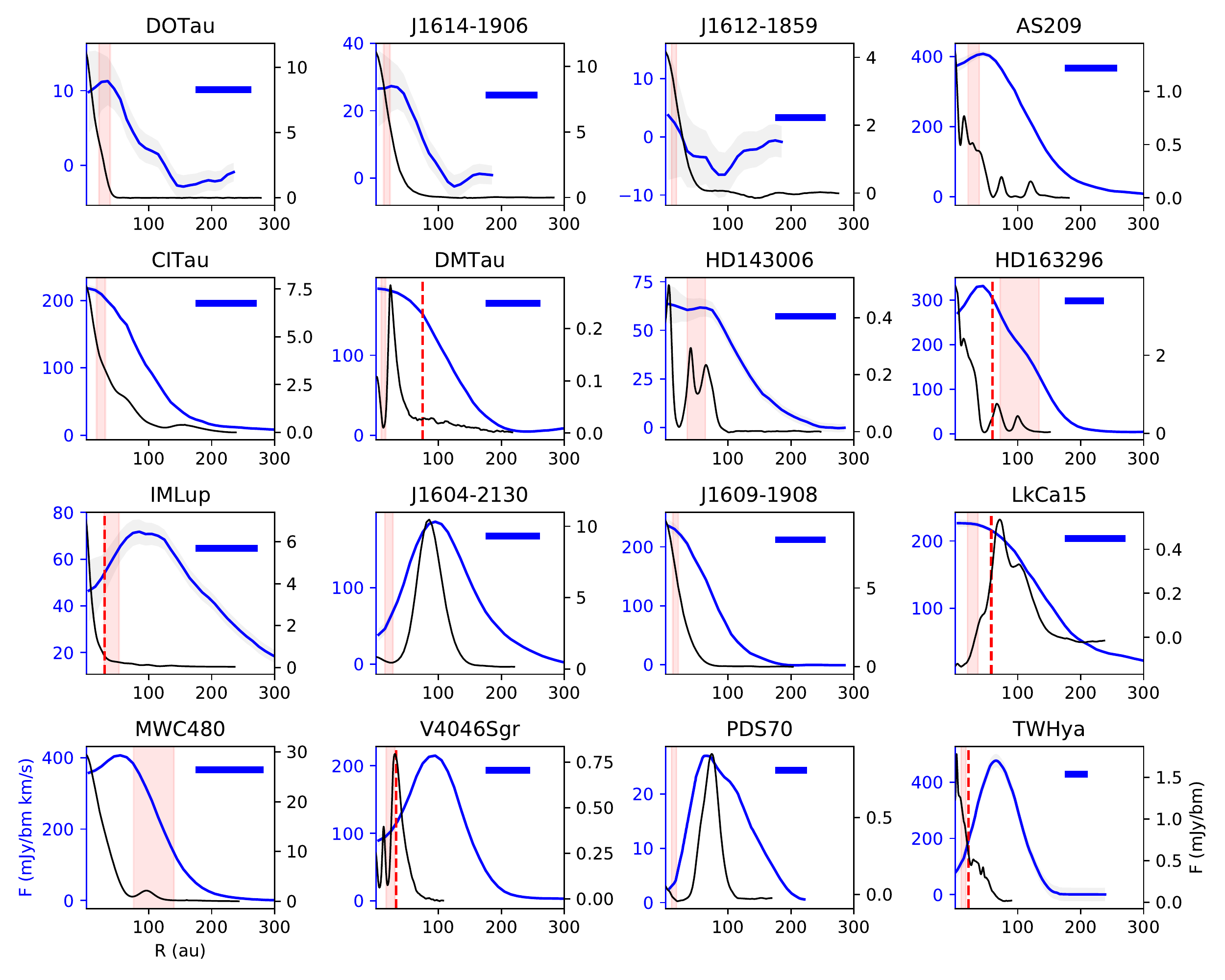}
    \caption{Radial C$_2$H profiles (in blue) of the disks in the Bergner sample plus TW Hya \citep{Bergin2016} and PDS~70 \citep{Facchini2021}. The grey indicates the uncertainty. The CO snowline (in the 22-30 K range) is overplotted as in Figure \ref{fig:profiles}. The high-resolution continuum profiles of Figure \ref{fig:profiles} are overplotted in black to show that the C$_2$H is always more extended than the continuum.}
    \label{fig:c2h_bergner}
\end{figure*}

Although the spatial resolution of the C$_2$H images is limited in current observations \citep{Bergner2019,Miotello2019}, the observed ring-like structures in their C$_2$H zero-moment maps and the well-resolved TW Hya disk \citep{Bergin2016} suggest that the C$_2$H emission originates outside the CO snowline (Figure \ref{fig:c2h_miotello} and \ref{fig:c2h_bergner}). Centrally peaked C$_2$H profiles such as seen in e.g. CI~Tau, J1609-1908 and LkCa~15 may just be the result of beam confusion and do not prove the presence of C$_2$H inside the CO snowline. The C$_2$H emission extends well beyond the radial dust extent, consistent with a gas disk much larger than a dust disk as icy dust pebbles settle and drift inwards, leaving the outer disk volatile-depleted. 

\section{Data reduction IRS 48 C$_2$H data}
\label{sec:irs48}
\citet{vanderMarel2021-irs48} presents a molecular line survey of Oph~IRS~48 from ALMA dataset 2017.1.00834.S for a number of H$_2$CO and CH$_3$OH transitions. The C$_2$H $N=4-3$,  $J=7/2-5/2$ doublet at 349.4 GHz is covered in these spectral windows as well, but not reported in this work. Inspection of the dataset reveals that this C$_2$H line is not detected. Considering the reported rms level of 1.2 mJy bm$^{-1}$ channel$^{-1}$ with a channel width of 1.6 km s$^{-1}$, a line width of $\sim$14 km s$^{-1}$ and a beam size of 0.63$\times$0.50" \citep{vanderMarel2021-irs48}, we compute a 3$\sigma$ upper limit on the C$_2$H flux density of 11 mJy bm$^{-1}$ km s$^{-1}$, or a scaled $T_{b,peak,100pc}$ limit of $<$1.1 K km s$^{-1}$. Such a low value of C$_2$H is consistent with the low CS/SO ratio reported by \citet{Booth2021-irs48}.

\section{ALMA archival continuum data 2017.1.01167.S}
For the continuum analysis of J1614-1906 and J1609-1908 we use ALMA archival data from ALMA Cycle 5 program 2017.1.01167.S (PI Sebastian Perez). These targets were observed in Band 6 for an integration time of 4.5 minutes each on January 18th, 2018. The dataset is reduced using CASA version 5.1.1 with the provided calibration scripts, with J1517-2422 as flux and bandpass calibrator and J1634-2058 as gain calibrator. We use the three continuum spectral windows (the fourth window was centered on the $^{12}$CO 2--1 line) with an average frequency of 240 GHz and a total bandwidth of 6 GHz. The continuum images are imaged using the CASA \texttt{tclean} task with uniform weighting, resulting in a beam size of 0.22$\times$0.20 and a rms noise of 0.08 mJy beam$^{-1}$. The total integrated flux is 21 and 17 mJy for J1609-1908 and J1614-1906, respectively. Both disks are marginally resolved and the CASA \texttt{uvmodelfit} indicates a radial extent of 0.14" (20 au) and 0.08" (11 au) for J1609 and J1614, respectively. Their disk morphology classification as compact disk is debatable due to the low resolution, and high-resolution imaging is required to properly assess these two disks. The images are presented in Figure \ref{fig:gallery}.

\end{document}